\documentclass[12pt]{article}
\usepackage[margin=1in]{geometry}
\usepackage{amsmath}
\usepackage{setspace}
\usepackage{hyperref}
\usepackage[round]{natbib}
\usepackage{graphicx}
\usepackage{epstopdf}

\title{Reference based multiple imputation - what is the right variance and how to estimate it}

\newcommand{\Var}{\mbox{Var}}

\author{Jonathan W. Bartlett \\ Department of Mathematical Sciences \\ University of Bath \\ UK}

\begin{document}
\maketitle

\large

\abstract{Reference based multiple imputation methods have become popular for handling missing data in randomised clinical trials. Rubin's variance estimator is well known to be biased compared to the reference based imputation estimator's true repeated sampling variance. Somewhat surprisingly given the increasingly popularity of these methods, there has been relatively little debate in the literature as to whether Rubin's variance estimator or alternative (smaller) variance estimators targeting the repeated sampling variance are more appropriate. We review the arguments made on both sides of this debate, and conclude that the repeated sampling variance is more appropriate. We review different approaches for estimating the frequentist variance, and suggest a recent proposal for combining bootstrapping with multiple imputation as a widely applicable general solution. At the same time, in light of the consequences of reference based assumptions for frequentist variance, we believe further scrutiny of these methods is warranted to determine whether the the strength of their assumptions are generally justifiable.}

\section{Introduction}
Reference or reference based imputation approaches have become popular for handling missing data in randomised trials  (\cite{Carpenter2013}). They have typically been used as a sensitivity analysis to a primary analysis which assumes data are missing at random (MAR). Unlike most missing not at random (MNAR) sensitivity analysis methods, which often require specification of the value of sensitivity parameters, reference based methods make assumptions which can be described somewhat more qualitatively by specifying the distribution of missing data in the active arm by reference to the distribution in the reference or control arm. Although reference based MI has since its inception tended to be used for sensitivity analyses for missing data, its core idea has been adopted more recently to handle intercurrent events for estimation of certain estimands  (\cite{mallinckrodt2019aligning}).

An important question when using reference based MI is how to estimate the variance of the resulting estimator. Conventionally the variance of MI estimators is obtained using Rubin's combination rules. However, as observed by \cite{seaman2014comment}, for reference based MI Rubin's variance estimator is biased upwards relative to the repeated sampling variance of the MI estimator. This bias is due to uncongeniality between the imputation and analysis models (\cite{Meng:1994}). Subsequently, a number of authors have discussed the merits of using either Rubin's variance estimator (\cite{carpenter2014response,cro2019information} or the frequentist variance (\cite{seaman2014comment,lu2014analytic,tang2017,white2020causal}), but no consensus has emerged.

In Section \ref{sec:refmi} we review reference based MI methods for continuous endpoints, the definition of congeniality, and why Rubin's variance estimator is biased upwards relative to the frequentist variance of the reference based point estimator of treatment effect. In Section \ref{sec:rightvar} we review arguments made in favour of both Rubin's variance estimator and the frequentist variance, concluding that if the assumptions made by reference based MI are employed, the frequentist variance is the right one. In Section \ref{sec:est_var} we review different approaches for estimating the frequentist variance of reference based estimators, and in Section \ref{sec:conclusions} give conclusions.

\section{Reference based multiple imputation and congeniality}
\label{sec:refmi}

In this section we review reference based multiple imputation methods and the congeniality issue. The approach was originally proposed in the context of a repeatedly measured continuous endpoint assuming a multivariate normal model (\cite{Carpenter2013}). Subsequently the idea has been extended to other endpoint types, including recurrent events (\cite{keene2014missing}) and survival times (\cite{atkinson2019reference}). To help make the following arguments regarding congeniality concrete yet (relatively) simple, we first review the jump to reference (J2R) approach for a repeatedly measured continuous endpoint, following \cite{Carpenter2013}.

\subsection{Jump to reference imputation}
We assume that $n$ patients are randomised to either reference (denoted $X_{i}=0$) or active treatment ($X_{i}=1$). Patients are scheduled to have the outcome measured at times $j=0,\dots,J$. The outcomes intended to be measured are thus $\mathbf Y_i=(Y_{i0},Y_{i1},Y_{i2},\dots,Y_{iJ})$. Often there may exist additional baseline covariates which are to be adjusted for, but to simplify the key points we do not include these.

For each patient, some measurements may be missing. We note that depending on the chosen estimand, the actual outcomes may be observed, but the potential outcomes of interest under the chosen estimand are missing. We assume that only monotone missingness occurs, although we note that implementations of J2R typically handle intermediate missingness using MAR imputation. Thus let $D_{i}$ denote the time of the last observation for patient $i$. A patient with complete follow-up thus has $D_{i}=J$. The MAR assumption says that the probability of each pattern of missingness occurring depends only on the data observed under that pattern (\cite{Tsiatis:2006}), which here means
\begin{eqnarray*}
P(D_{i}=j | Y_{i0},Y_{i1},\dots,Y_{iJ},X_{i}) = P(D_{i}=j | Y_{i0},Y_{i1},\dots,Y_{ij},X_{i}),
\end{eqnarray*}
i.e. that the (marginal) probability of dropping out immediately after time $j$ does not (statistically) depend on $Y_{j+1},\dots,Y_{J}$, conditional on the outcome measurements obtained up to and including time $j$. Alternatively, as described by \cite{Daniels2008} it can be equivalently stated as
\begin{eqnarray*}
P(D_{i}=j | D_{i}\geq j, Y_{i0},Y_{i1},\dots,Y_{iJ},X_{i}) = P(D_{i}=j | D_{i} \geq j, Y_{i0},Y_{i1},\dots,Y_{ij},X_{i}),
\end{eqnarray*}
i.e. that among those who have not yet `dropped out' at time $j$, the probability that they drop out before time $j+1$ does not depend on the outcomes $Y_{j+1},\dots,Y_{J}$, conditional on the outcomes measured through to time $j$.

In the reference arm we assume that $\mathbf Y_i=(Y_{i0},Y_{i1},\dots,Y_{iJ})$ is distributed multivariate normal with a distinct mean at each follow-up time and an unstructured covariance matrix, and this model is fitted to the observed data using maximum likelihood, assuming MAR. The same model is separately fitted to the data in the active arm, with distinct parameters. To generate an imputed dataset, from each model a posterior draw is taken of the respective model parameters. Following \cite{Carpenter2013}, we let $\boldsymbol \mu_{r}=(\mu_{r,0},\dots,\mu_{r,J})^{T}$ and unstructured covariance matrix $\boldsymbol \Sigma_{r}$ denote the resulting posterior draws of the mean and covariance matrix in the reference arm, and $\boldsymbol \mu_{a}$ and $\boldsymbol \Sigma_{a}$ for the corresponding parameters in the active arm, with
\begin{align*}
\boldsymbol \Sigma_{r} &= \begin{bmatrix} \boldsymbol R_{11} & \boldsymbol R_{12} \\ \boldsymbol R_{21} & \boldsymbol R_{22} \end{bmatrix} \\
\mathbf \Sigma_{a} &= \begin{bmatrix} \boldsymbol A_{11} & \boldsymbol A_{12} \\ \boldsymbol A_{21} & \boldsymbol A_{22} \end{bmatrix}
\end{align*}

Consider a patient with observed values $Y_{i0},Y_{i1},\dots,Y_{iD_{i}}$ and missing values $Y_{i(D_{i}+1)},\dots,Y_{iJ}$. If they are in the reference arm, their missing values are imputed from the conditional distribution implied by the assumed model under MAR and the reference arm model parameter posterior draws $\boldsymbol \mu_{r}$ and $\boldsymbol \Sigma_{r}$. If the patient is in the active treatment arm they are imputed using the multivariate normal conditional distribution implied by assuming that their full data vector has marginal mean equal to
\begin{align*}
\boldsymbol{\tilde{\mu}_{i}} = (\mu_{a,0},\dots,\mu_{a,D_{i}},\mu_{r,D_{i}+1},\dots,\mu_{r,J})^{T}.
\end{align*}
and variance covariance equal to
\begin{align*}
\boldsymbol{\tilde{\Sigma}} = \begin{bmatrix} \boldsymbol{\tilde{\Sigma}_{11}} & \boldsymbol{\tilde{\Sigma}_{12}} \\ \boldsymbol{\tilde{\Sigma}_{21}} & \boldsymbol{\tilde{\Sigma}_{22}} \end{bmatrix}
\end{align*}
where
\begin{align*}
\boldsymbol{\tilde{\Sigma}_{11}} &= \boldsymbol A_{11} \\
\boldsymbol{\tilde{\Sigma}_{21}} &= \boldsymbol R_{21} \boldsymbol R^{-1}_{11} \boldsymbol A_{11} \\
\boldsymbol{\tilde{\Sigma}_{22}}&= \boldsymbol R_{22} - \boldsymbol R_{21} \boldsymbol R^{-1}_{11} (\boldsymbol R_{11}-\boldsymbol A_{11}) \boldsymbol R^{-1}_{11} \boldsymbol R_{12} \\
\end{align*}
The latter values are those that ensure that the sub-matrix of $\boldsymbol{\tilde{\Sigma}}$ corresponding to the observed measurements matches that in the active arm and the conditional covariance matrix of the missing components given the observed matches that in the reference arm.

The preceding describes the J2R approach. \cite{Carpenter2013} also proposed a number of other variants, including last mean carried forward (LMCF), copy increments in reference and copy reference.

\subsection{Uncongeniality of reference based imputation}
\cite{seaman2014comment} and \cite{lu2014analytic} both noted that reference based imputation approaches were uncongenial with what would be the standard analysis of the resulting imputed data, namely a linear regression model of the final time point outcome, with randomised treatment as covariate (plus other baseline covariates typically). As such, Rubin's variance estimator is not unbiased (even asymptotically) for the true repeated sampling variance of the estimator of treatment effect obtained after using reference based MI to impute missing values.

To investigate why reference based MI leads to uncongeniality, we first review the definition of congeniality (\cite{Meng:1994,xie2017dissecting}). We let $\boldsymbol{Z_{\text{com}}}=\{(X_{i},D_{i},Y_{i0},\dots,Y_{iJ}); i=1,\dots,n\}$ denote the complete data, $\boldsymbol{Z_{\text{mis}}}$ denote the missing component, and $\boldsymbol{Z_{\text{obs}}}$ denote the observed component. Let $\theta$ denote the parameter of interest, which here is the difference in mean outcomes between randomised groups at the final time point, $\theta=E(Y_{J}|X=1)-E(Y_{J}|X=0)$. In the context considered by \cite{Meng:1994}, the imputer and analyst in general are distinct entities, whereas in the present context of clinical trials they are the same person. In the absence of any other baseline covariates, the analysis model is a linear regression model for $Y_{J}$ with $Y_{0}$ and $X$ as covariates. In Meng's terminology, the `analyst's complete data procedure' is the ordinary least squares estimator of the coefficient of $X$ in this regression, which we denote $\hat{\theta}^{A}(\boldsymbol{\tilde{Z}_{\text{com}}})$, where A stands for analyst and $\boldsymbol{\tilde{Z}_{\text{com}}}$ is an arbitrary complete dataset. The analyst's variance estimator, $W^{A}(\boldsymbol{\tilde{Z}_{\text{com}}})$, is the standard model based variance estimator from linear regression for the coefficient of randomised treatment group $X$.

Following \cite{xie2017dissecting} and \cite{Bartlett:Hughes2020}, the imputation model and the analyst's complete data procedure are said to be congenial if there exists a unifying Bayesian model (referred to by $IA$) which embeds the imputer's imputation model and the analyst's complete data procedure, in the sense that
\begin{enumerate}
\item For all possible complete datasets $\boldsymbol{\tilde{Z}_{\text{com}}}$,
 	\begin{eqnarray} 
	\hat{\theta}^{A}(\boldsymbol{\tilde{Z}_{\text{com}}}) = E^{IA}(\theta|\boldsymbol{\tilde{Z}_{\text{com}}}) \text{ and } W^{A}(\boldsymbol{\tilde{Z}_{\text{com}}})=\Var^{IA}(\theta|\boldsymbol{\tilde{Z}_{\text{com}}})
	\label{congenial1}
	\end{eqnarray}
where $E^{IA}$ and $\Var^{IA}$ denote posterior expectation and variance with respect to the embedding Bayesian model;
\item For all possible $\boldsymbol{\boldsymbol{\tilde{Z}_{\text{mis}}}}$,
	\begin{eqnarray}
	f^{I}(\boldsymbol{\tilde{Z}_{\text{mis}}}|\boldsymbol{Z_{\text{obs}}}) = f^{IA}(\boldsymbol{\tilde{Z}_{\text{mis}}}|\boldsymbol{Z_{\text{obs}}})
		\label{congenial2}
	\end{eqnarray}
	where $f^{I}(\boldsymbol{\tilde{Z}_{\text{mis}}}|\boldsymbol{Z_{\text{obs}}})$ denotes the predictive distribution for the missing data used by the imputation model and $f^{IA}(\boldsymbol{\tilde{Z}_{\text{mis}}}|\boldsymbol{Z_{\text{obs}}})$ is the predictive distribution for the missing data given the observed data under the embedding Bayesian model.
\end{enumerate}

To see more clearly why reference based MI leads to uncongeniality, following \cite{seaman2014comment} and \cite{carpenter2014response} we consider an unrealistic but instructive situation where we omit the baseline measurement $Y_{0}$ and set $J=1$. Thus now $D=0$ indicates that $Y_1$ is missing and $D=1$ indicates $Y_1$ is observed. With no baseline measurement, the analyst's complete data procedure reduces to calculating the difference in mean of $Y_{1}$ between those randomised to active ($X=1$) and those randomised to reference ($X=0$):
\begin{equation}
\hat{\theta}^{A}(\boldsymbol{\tilde{Z}_{\text{com}}}) = \frac{\sum^{n}_{i=1} Y_{i1} X_{i}}{\sum^{n}_{i=1} X_{i}} - \frac{\sum^{n}_{i=1} Y_{i1}(1-X_{i})}{\sum^{n}_{i=1} 1-X_{i}}
\label{completeDataMeanDiff}
\end{equation}
The J2R imputation model in this highly simplified case assumes that
\begin{align}
Y_{1} | D=0, X=0 & \sim N(\mu_{r}, \sigma^{2}_{r}) \nonumber \\
Y_{1} | D=1, X=0 & \sim N(\mu_{r}, \sigma^{2}_{r}) \nonumber \\
Y_{1} | D=0, X=1 & \sim N(\mu_{r}, \sigma^{2}_{r}) \nonumber \\
Y_{1} | D=1, X=1 & \sim N(\mu_{a}, \sigma^{2}_{a}),
\label{j2rImpModelAss}
\end{align}
such that all outcomes have mean $\mu_{r}$ except those in the active arm with $D=1$, who have mean $\mu_{a}$. The J2R imputation model for the missing data can be embedded in a model for the complete data with $f(Y,D,X)=f(Y|D,X)P(D|X)P(X)$ in which $f(Y|D,X)$ is given by the normal models in equation \eqref{j2rImpModelAss} and $P(D=0|X=x)=\pi_{x}, \; x=0,1$, so that $E(D|X=x)=1-\pi_{x}$. We do not specify $P(X)$ (although we know its distribution from the randomisation scheme), but rather perform inference conditional on $X$.

Given an arbitrary complete dataset $\boldsymbol{\tilde{Z}_{\text{com}}}$, under the model embedding J2R imputation the MLE of $\mu_{r}$ is the mean outcome combining the reference arm patients with those in the active arm with $D=0$, which can be expressed as:
\begin{align*}
\hat{\mu}^{\text{com}}_{r} = \frac{\sum^{n}_{i=1} Y_{i1} (1-D_i X_i) }{\sum^{n}_{i=1} 1-D_i X_i},
\end{align*}
and the MLE of $\mu_{a}$ is the mean outcome in those with $X=1$ and $D=1$:
\begin{align*}
\hat{\mu}_{a} = \frac{\sum^{n}_{i=1} Y_{i1} D_i X_i }{\sum^{n}_{i=1} D_i X_i}
\end{align*}
Lastly, the MLE of $\pi_{x}$ for $x=0,1$ is simply the sample proportion with $D=0$ in the $X=0$ and $X=1$ treatment groups. Then under the embedding model we have that
\begin{align}
\theta &= E(Y|X=1)-E(Y|X=0) \nonumber \\
&= E\left[E(Y|X=1,D) | X=1 \right] - E\left[E(Y|X=0,D) | X=0 \right] \nonumber \\
&= E\left[\mu_{r} + (\mu_{a}-\mu_{r}) D | X=1 \right] - E(\mu_{r} | X=0) \nonumber \\
&= \mu_{r} + (\mu_{a}-\mu_{r}) E(D | X=1) - \mu_{r} \nonumber \\
&= (\mu_{a}-\mu_{r}) (1-\pi_{1}) \label{j2r_theta} 
\end{align}
The MLE of $\theta$ given complete data under the embedding model follows from this expression under the invariance property of MLE. Morever, for large $n$ the posterior mean under the embedding model is (essentially) equal to the MLE, so that for large $n$ we have
\begin{align}
E^{IA}(\theta|\boldsymbol{\tilde{Z}_{\text{com}}}) &= (\hat{\mu}_{a}-\hat{\mu}^{\text{com}}_{r}) (1-\hat{\pi}_{1}) \nonumber \\
&= \left( \frac{\sum^{n}_{i=1} Y_{i1} D_i X_i }{\sum^{n}_{i=1} D_i X_i} -  \frac{\sum^{n}_{i=1} Y_{i1} (1-D_i X_i) }{\sum^{n}_{i=1} 1-D_i X_i} \right) \frac{\sum^{n}_{i=1} D_{i}X_{i}}{\sum^{n}_{i=1} X_{i}}
\label{completeDataMLE}
\end{align}
which, unlike the analyst's complete data estimator, uses $D$, and is not equal to the analyst's complete data estimator $\hat{\theta}^{A}(\boldsymbol{\tilde{Z}_{\text{com}}})$. Indeed, suppose that in the active arm virtually all patients were missing their outcomes, such that $\hat{\pi}_{1} \approx 1$. In this case $E^{IA}(\theta|\boldsymbol{\tilde{Z}_{\text{com}}}) \approx 0$, whereas the analyst's complete data estimator is not (in general). Thus the first part of the first condition in the congeniality definition is not satisfied.

Despite the fact that the analyst's complete data estimator does not in general match the complete data posterior mean under the embedding model, the J2R MI estimator for $\theta$ which uses the analyst's complete data estimator \textit{is} equivalent (with an infinite number of imputations) to the Bayesian posterior mean under the embedding model. To see this, following \cite{seaman2014comment} and \cite{carpenter2014response}, consider the MLE/posterior mean of $\theta$ given the observed data under the model embedding J2R imputation. The only change moving from the complete data to the observed data is that the MLE of $\mu_r$ is now based only on reference arm patients with $D=1$, so that
\begin{align}
E^{IA}(\theta|\boldsymbol{Z_{\text{obs}}}) &= (\hat{\mu}_{a}-\hat{\mu}^{\text{obs}}_{r}) (1-\hat{\pi}_{1} )
\label{j2rObsDataMLE}
\end{align}
As the number of imputations goes to infinity, and provided $n$ is sufficiently large for the priors to have essentially no impact, under J2R MI the active group mean converges to $(1-\hat{\pi}_{1}) \hat{\mu}_{a} + \hat{\pi}_{1} \hat{\mu}^{\text{obs}}_{r}$, whereas the control group mean converges to $\hat{\mu}^{\text{obs}}_{r}$. Thus the J2R MI estimator of $\theta$ converges to (as the number of imputations tends to infinity)
\begin{align}
(1-\hat{\pi}_{1}) \hat{\mu}_{a} + \hat{\pi}_{1} \hat{\mu}^{\text{obs}}_{r} - \hat{\mu}^{\text{obs}}_{r} = (\hat{\mu}_{a}-\hat{\mu}^{\text{obs}}_{r}) (1-\hat{\pi}_{1}) = E^{IA}(\theta|\boldsymbol{Z_{\text{obs}}})
\label{j2rMIEst}
\end{align}

Turning to the variance, we must check whether the analyst's complete data variance $W^{A}(\boldsymbol{\tilde{Z}_{\text{com}}})$ matches the posterior variance given complete data under the embedding model. The analyst's complete data variance estimator could either assuming equal variances for $Y$ in the two groups (i.e. the standard t-test) or could use the variance estimator which relaxes this assumption, by estimating the variance separately in each group. Like \cite{carpenter2014response}, we will assume the analyst does the latter, so that
\begin{align}
W^{A}(\boldsymbol{\tilde{Z}_{\text{com}}}) = \frac{\widehat{\text{Var}}(Y|X=1)}{n_a} +  \frac{\widehat{\text{Var}}(Y|X=0)}{n_r}
\label{analystCompleteDataVar}
\end{align}
where $n_a$ and $n_r$ denote the number randomised to active and control, and $\widehat{\text{Var}}(Y|X=1)$ and $\widehat{\text{Var}}(Y|X=0)$ are the sample variances in each treatment group. Assuming the J2R assumptions (equation \eqref{j2rImpModelAss}), we have $\Var(Y|X=0)=\sigma^{2}_{r}$. For $\Var(Y|X=1)$, we can use the law of total variance to give that
\begin{align}
\Var(Y|X=1) &= \Var\left[E(Y|X=1,D) | X=1\right] + E \left[ \Var(Y|X=1,D) | X=1\right] \nonumber \\
&= \Var\left[\mu_{r} + (\mu_{a}-\mu_{r}) D | X=1\right] + E \left[ \sigma^{2}_{a} D + \sigma^{2}_{r}(1-D) | X=1\right] \nonumber \\
&= (\mu_a-\mu_r)^{2} \pi_{1}(1-\pi_{1}) + \sigma^{2}_{a}(1-\pi_{1}) + \sigma^{2}_{r} \pi_{1}
\label{varYX1}
\end{align}
For the complete data posterior variance under the embedding model, again suppose $n$ is large so that this matches the MLE estimated variance based on the observed information matrix. Some algebra shows that this is equal to
\begin{align}
\Var^{IA}(\theta|\boldsymbol{\tilde{Z}_{\text{com}}}) =
(1-\hat{\pi}_{1})\left[\frac{\hat{\sigma}^{2}_{r}(1-\hat{\pi}_{1})}{n_{r}+n_{a}\hat{\pi}_{1}} + \frac{\hat{\sigma}^{2}_{a}}{n_a} + \frac{(\hat{\mu}^{\text{com}}_{r}-\hat{\mu}_{a})^{2} \hat{\pi}_{1}}{n_a} \right]
\label{mleCompleteDataVar}
\end{align}
where $\hat{\sigma}^{2}_{a}$ is the estimated variance of $Y$ from those with $X=1$ and $D=1$ and $\hat{\sigma}^{2}_{r}$ is the estimated variance from the remaining patients (i.e. $X=0$, or $X=1$ and $D=0$). Equations \eqref{analystCompleteDataVar} and \eqref{mleCompleteDataVar} are not the same, as required for the second part of the first condition in the definition of congeniality. For example, consider again the case that almost all patients in the active arm have missing data, such that $\hat{\pi}_{1} \approx 1$. Then from equation \eqref{completeDataMLE} the complete data posterior mean is approximately zero and from equation \eqref{mleCompleteDataVar} its estimated variance is also approximately zero. In contrast, if $\pi \approx 1$, from equation \eqref{varYX1} $\Var(Y|X=1) \approx \sigma^{2}_{r}$, and the analyst's complete data variance estimator of equation \eqref{analystCompleteDataVar} will (on average) estimate $\sigma^{2}_{r}(n^{-1}_{a} + n^{-1}_{r})$, i.e. greater than zero. Thus the second part of the first condition in the congeniality definition is also not satisfied. 

Rubin's rules variance estimator is based on decomposing the posterior variance of $\theta$ under the embedding model as 
\begin{align*}
\Var&^{IA}(\theta | \boldsymbol{Z_{\text{obs}}}) = E^{IA}\left[\Var^{IA}(\theta|\boldsymbol{\tilde{Z}_{\text{com}}}) | \boldsymbol{Z_{\text{obs}}}\right] +  \Var^{IA}\left[E^{IA}(\theta|\boldsymbol{\tilde{Z}_{\text{com}}}) | \boldsymbol{Z_{\text{obs}}}\right]
\end{align*}
Rubin's rules approximates the first part, the within-imputation variance, by substituting $W^{A}(\boldsymbol{\tilde{Z}_{\text{com}}})$ for $\Var^{IA}(\theta|\boldsymbol{\tilde{Z}_{\text{com}}})$. Since as we have seen $W^{A}(\boldsymbol{\tilde{Z}_{\text{com}}})$ is too large, this component will be biased upwards. Rubin's rules approximates the second part, the between-imputation variance, by subtituting $\hat{\theta}^{A}(\boldsymbol{\tilde{Z}_{\text{com}}})$ for $E^{IA}(\theta|\boldsymbol{\tilde{Z}_{\text{com}}})$. Consider again the case where $\hat{\pi}_{1} \approx 1$. As we have noted previously, from equation \eqref{completeDataMLE}, $E^{IA}(\theta|\boldsymbol{\tilde{Z}_{\text{com}}}) \approx 0$, and so $\Var^{IA}\left[E^{IA}(\theta|\boldsymbol{\tilde{Z}_{\text{com}}}) | \boldsymbol{Z_{\text{obs}}}\right] \approx 0$. In contrast, the value of $\hat{\theta}^{A}(\boldsymbol{\tilde{Z}_{\text{com}}})$ will vary across imputations, so that the estimated between-imputation variance will be larger than zero.

In conclusion, the observed data posterior mean of $\theta$ under the embedding model essentially matches the J2R MI estimator of $\theta$. Rubin's rules variance estimator is however larger than the observed data posterior variance under the embedding model. Assuming the embedding model is correct, the latter will (asymptotically) estimate the true frequentist variance of the point estimator correctly, and thus we conclude Rubin's variance estimator is biased upwards compared to the true frequentist variance of the point estimator.

Regarding other variants of reference based MI, positive bias in Rubin's variance estimator for a copy reference type approach was shown through simulation by \cite{lu2014analytic} and \cite{tang2017} derived analytical expressions for the frequentist bias in Rubin's variance estimator for J2R and copy reference MI. \cite{gao2017inference} showed by simulation upward bias in Rubin's variance estimator in the case of repeated binary outcomes.

\section{What is the right variance for reference based multiple imputation?}
\label{sec:rightvar}

As described in the previous section, for reference based MI, Rubin's variance estimator is biased relative to the true repeated sampling variance of the point estimator of the treatment effect. How large the estimated variance of the treatment effect estimator should be is obviously important, since it affects the advertised precision of the estimated treatment effect and consequently type 1 error and power to detect an effect. As expected given the upward (frequentist) bias in Rubin's variance estimator, simulation studies have shown that use of referenced based MI with Rubin's rules leads to conservative type 1 error control under the null and the potential for substantial power loss (compared to using the frequentist variance) under the alternative (\cite{gao2017inference,lu2014analytic,tang2017}).

\cite{carpenter2014response} argued that the frequentist repeated sampling variance is inappropriate in the context of using reference based MI as a sensitivity analysis to a primary analysis which handles the missing data under a `baseline' assumption, e.g. MAR. They a proposed principle that for missing data sensitivity analyses, the variance should be no lower (on average) than the complete data variance estimator, and they showed that reference based MI with the frequentist variance violates this principle.

\cite{cro2019information} developed this principle further, considering a trial in which missing data are first handled using a `primary' set of assumptions about missingness and second handled using an alternative `sensitivity' set of assumptions. They defined an information anchored sensitivity analysis (e.g. an analysis using J2R) as one in which the relative loss in information about $\theta$ caused by missing data is the same as the loss in the primary analysis. \cite{cro2019information} argued that in the context of trials, information anchored sensitivity analyses seem appropriate because, relative to the primary analysis assumptions, they are neither adding or removing information. They showed that reference based MI such as J2R are approximately information anchored when Rubin's variance estimator is used for inferences, whereas the repeated sampling variance is information positive - information is being added relative to an MAR based analysis.

In contrast, others have argued that the repeated sampling variance may be more appropriate. \cite{white2020causal} and \cite{gao2017inference} pointed out that using reference based MI with Rubin's rules leads to type 1 error rates which are too small under the null and a loss of power under the alternative, and as such when used for the primary analysis, the frequentist variance may be preferable.

Consider the use of reference based MI as a primary analysis estimator of treatment effect. \cite{cro2019information} quite reasonably point out that it seems very counterintuitive to use a method which apparently is able to make more precise inferences the more data is lost. As we have seen however, this is a logical consequence of the strength of the assumption made by reference based MI methods. If such behaviour is viewed as undesirable, and we believe that in many settings it may be, we believe the correct response is to conclude that the assumptions made by the reference based approach are inappropriate, rather than to assign blame to a variance estimator. Indeed, the uncongeniality issue here is caused by the fact we are happy to make a (strong) assumption in the imputation model but not in the analysis. If we truly believe in the assumption or at least want to perform an analysis supposing it is true, we should use it throughout our analysis (i.e. at both imputation and analysis stages). If we do not believe in it, or feel it is too strong, we should not use it.

Turning next to the use of referenced based MI for sensitivity analyses, we agree with \cite{cro2019information} that ensuring that sensitivity analyses do not inject or take away information (precision) relative to a primary set of missing data assumptions seems like a reasonable principle to adhere to. However, when considering this statement we believe it is critical to be careful about the precise meaning of `information'. \cite{cro2019information} implicitly that the view that information corresponds to \textit{estimates} of the variance of point estimators, rather than the true repeated sampling variance of the point estimators. Relative to an MAR analysis, reference based MI estimators such as J2R \textit{do} inject information when information is judged in terms of the true repeated sampling variance of the estimator. Using reference based MI with Rubin's rules to estimate the variance in our view amounts to pretending reference based assumptions about missing data are information anchored to an MAR analysis, when in actual fact they are information positive.

In summary, we believe that under a frequentist inference paradigm, information (precision) should be judged in terms of the true repeated sampling variance of estimators. If one wishes to perform information anchored sensitivity analyses, we believe the correct solution is to construct missing data assumptions which differ to those made by the primary analysis but which genuinely neither add nor remove information, with information being judged in terms of the estimator's true repeated sampling variance. \cite{cro2019information} propose one possible route to this - adding additional random noise to the reference based MI estimator so that its true repeated sampling variance matches the primary analysis method's variance. A drawback of this approach is that it would then seem difficult to readily communicate the totality of the missing data assumptions made by such a `added noise' reference based MI estimator. We thus believe further research is warranted to develop sensitivity analysis methods which are information anchored in the sense described above but which like referenced based methods can be relatively easily communicated.

\section{Estimating the repeated sampling variance}
\label{sec:est_var}
In this section we review methods for estimating the frequentist variance of reference based MI estimators, considering their relative advantages and disadvantages.

\subsection{Analytical variance estimators}
A number of authors have developed analytical variance estimators for reference based estimators for various endpoint types. For a continuous endpoint \cite{lu2014analytic} developed a maximum likelihood estimator with delta method variance under a copy reference assumption. \cite{tang2015short} derived equivalent matrix versions of \cite{lu2014analytic}'s copy reference estimator and accompanying delta method variance estimator. \cite{tang2017} derived analytical variance estimators for J2R and copy reference methods. \cite{gao2017inference} applied the general theory developed by \cite{Robins/Wang:2000} to derive analytical variance estimators for reference based MI in the setting of repeated binary data. In all the preceding papers the analytical variance estimators show good type 1 error control under the null in simulations, and improved power under the alternative compared to using Rubin's rules.

Analytical variance estimators have the major advantage of being computationally fast. Their drawback however is that they must be derived specifically for each case and implemented in software. Moreover, as noted by \cite{gao2017inference}, there are situations where it may be difficult to derive such variance estimators, for example when intermediate missing values are imputed assuming MAR in a first stage followed by use of reference based imputation, or perhaps when different imputation strategies are used for different types of intercurrent event.

\subsection{Congenial Bayesian approach}
An alternative approach is to perform (congenial) Bayesian inference for the treatment effect under a model which embeds the reference based assumptions. This approach was developed by \cite{lu2014analytic} and \cite{liu2016analysis}. Since it is relatively straightforward to obtain posterior draws of the MAR MMRM models using existing software, provided one can express the treatment effect (under the assumed reference based assumption) as a function of the model parameters (equation \eqref{j2r_theta} being an example), one can obtain posterior draws of the treatment effect under this assumption by simply applying this function to the posterior draws of the MMRM model parameters. For large $n$ this approach results in accurate frequentist inferences, provided the assumed model is correct. We emphasize that here congeniality is not an issue here because one constructs the expression for the treatment effect under the assumed reference based assumpion - there is no uncongenial analyst complete data estimator as there is with the reference based MI approach.

A possible drawback with this approach however is that, like analytical variance estimators, expressing the treatment effect as a function of the MMRM model parameters may become complex and setup specific, for example if one wanted to make a variety of different imputation assumptions to handle different types of intercurrent events.

\subsection{Bootstrap variance estimators}
\label{sec:est_var_boot}
As noted previously, maximum likelihood type analytical variance estimators and the congenial Bayesian approach require problem specific derivations and implementations. An alternative which avoids this, at the expense of computational cost, is to use bootstrapping. \cite{gao2017inference} proposed applying nonparametric bootstrapping to reference based MI estimators in the context of control based MI for repeated binary endpoints. Simulations showed it gave type 1 error control close to the nominal level under the null and superior power to using Rubin's rules under the alternative. \cite{gao2017control} and \cite{diao2020efficient} similarly used the same bootstrapping approach for reference based MI estimators for recurrent event data, while \cite{quan2018considerations} examined its use for reference based MI with continuous endpoints. \cite{zhang2020likelihood} examined the performance of bootstrapping when used with a return to baseline type MI approach. We emphasize that under uncongeniality it is critical for the bootstrapping to be applied first, followed by multiple imputation. Approaches based on first multiply imputing missing data and then bootstrapping are not generally valid under uncongeniality (\cite{Bartlett:Hughes2020}).

While there is now extensive empirical evidence showing that the nonparametric bootstrap can deliver accurate frequentist inference when used with reference based MI estimators, its major drawback is its large computational cost. Whereas standard MI is often performed using a relative small number of imputations, bootstrapping requires a much larger number of bootstraps to give accurate inferences. This high computational cost can first be partly mitigated by noting that once bootstrapping is used for inference, there is no need for the multiple imputation to be `proper' (\cite{vonHippelbartlett2021}). This means that it suffices to generate each imputed dataset conditional on efficient estimates (e.g. MLE) of the imputation model parameters. To implement this one needs to make a generally minor modification to existing software, by skipping the step in the algorithm that performs the posterior draw - whether this be an analytical posterior draw or one based on Markov Chain Monte-Carlo (MCMC) methods. In the case where MCMC methods are used to obtain posterior draws, removing this step is a major advantage because first it removes the computational cost of running the chains, and second because one does not need to be concerned with how many iterations are required for stationarity and independence of draws to be achieved.

One might be tempted to reduce computational cost by reducing the number of imputations performed on each bootstrap sample. This however increases the Monte-Carlo noise in the point estimate of treatment effect estimator, leading to a somewhat less precise effect estimate and wider confidence intervals than are necessary (\cite{Bartlett:Hughes2020}).

An alternative bootstrap approach which overcomes this issue was proposed by \cite{vonHippelbartlett2021}. Their approach performs a small (e.g. two) number of imputations of each bootstrap. The point estimator is taken as the average of estimates across all bootstraps and imputations. To estimate the variance of this estimator, they fit a simple random intercepts model to estimate the between bootstrap and within-bootstrap (between imputation) variances. \cite{Bartlett:Hughes2020} demonstrate that the approach of \cite{vonHippelbartlett2021} provides efficient frequentist valid inferences under uncongeniality yet is substantially quicker to run compared to applying standard non-parametric bootstrapping to an MI estimator which uses a large number of imputations.

\section{Conclusions}
\label{sec:conclusions}
It has been known for almost 10 years that Rubin's variance estimator is biased upwards relative to the true repeated sampling variance of reference based estimators of treatment effects in randomised trials, but there remains no settled view on what is the right variance to use. Given the increasing use of reference based MI in trials, this is not particularly satisfactory. We have argued that the frequentist variance is the correct variance for referenced based estimators. If the behaviour of the frequentist variance does not seem appropriate to the analyst, for example because it decreases as the amount of missing data in the active arm increases, then our view is that this means the analyst does not really belief the assumptions made by the reference based approach.

In the context of performing missing data sensitivity analyses, the proposed principle that sensitivity analyses should be information neutral, or anchored, seems eminently sensible. However, we believe that provided we are operating under the frequentist paradigm, information here must be judged in terms of repeated sampling variance. In our view, using reference based MI with Rubin's variance estimator amounts to using a point estimator that is not information anchored but then using a variance estimator that falsely suggests it is information anchored. Further research is warranted to develop new sensitivity analyses which retain the attractive features of a reference based type approach, where assumptions are structured qualitatively, rather than quantitatively, but which are information anchored in the frequentist variance sense.

Historically referenced based estimators have tended to be mostly used as sensitivity analyses to a primary analysis which adopts the MAR assumption. However, such methods, and combinations of them are being increasingly used to develop estimators which might be used as the primary analysis method of trials (e.g. \cite{darken2020}). In this context it is clearly important to assess whether the strong assumptions potentially made by the missing data assumptions of such approaches are justifiable, particularly in light of what they imply for the repeated sampling variance of the treatment effect estimator.

We have suggested that a particular way of combining bootstrapping with MI can be used to obtain estimates of frequentist variance of referenced based MI estimators. Because Rubin's rules are no longer used, the imputation process does not need to be `proper', and imputation can instead be performed conditional on maximum likelihood estimates. As described by \cite{vonHippelbartlett2021}, to implement this in software should in most cases require relatively small changes, since the part that performs the draw from the posterior distribution (e.g. via MCMC sampling) can simply be skipped. In R, J2R MI is implemented for continuous endpoints using this approach in the \texttt{mlmi} package, while the \texttt{RefBasedMI} package\footnote{\url{https://github.com/UCL/RefbasedMI}}, currently under development, has an option that allows the user to impute conditional on the MLE for a much wider range of reference based assumptions for continuous endpoints. The R package \texttt{dejaVu} implements referenecd based MI for recurrent event data, as proposed by \cite{keene2014missing}, and also includes an option to impute conditional on the maximum likelihood estimates. The calculations to implement the bootstrap/MI approach proposed by \cite{vonHippelbartlett2021} are relatively simple, but for R users they are implemented in the \texttt{bootImpute} package.

\section*{Acknowledgements}
This work was supported by MRC Grant MR/T023953/1.


\end{document}